\documentclass[twocolumn,twocolappendix,tighten]{aastex631} 
\usepackage[caption=false]{subfig}
\usepackage{hyperref}
\shorttitle{Fiber Spillover in Broad Line AGNs}
\shortauthors{Pfeifle et al.}
\graphicspath{{./}{figures/}}

\begin{document}

\title{The Messy Nature of Fiber Spectra: Star-Quasar Pairs Masquerading as Dual Type~1 AGNs}


\correspondingauthor{Ryan W. Pfeifle}
\email{ryan.w.pfeifle@nasa.gov}

\author[0000-0001-8640-8522]{Ryan W. Pfeifle}
\altaffiliation{NASA Postdoctoral Program Fellow}
\affiliation{X-ray Astrophysics Laboratory, NASA Goddard Space Flight Center, Code 662, Greenbelt, MD 20771, USA}
\affiliation{Oak Ridge Associated Universities, NASA NPP Program, Oak Ridge, TN 37831, USA}

\author[0000-0003-2283-2185]{Barry Rothberg}
\affiliation{Large Binocular Telescope Observatory, University of Arizona, 933 N. Cherry Ave, Tucson, USA}
\affiliation{Department of Physics and Astronomy, George Mason University, 4400 University Drive, MSN 3F3, Fairfax, VA 22030, USA}

\author{Kimberly A. Weaver}
\affiliation{X-ray Astrophysics Laboratory, NASA Goddard Space Flight Center, Code 662, Greenbelt, MD 20771, USA}

\author[0000-0003-3432-2094]{Remington O. Sexton}
\affiliation{U.S. Naval Observatory, 3450 Massachusetts Avenue NW, Washington, DC 20392, USA}
\affiliation{Department of Physics and Astronomy, George Mason University, 4400 University Drive, MSN 3F3, Fairfax, VA 22030, USA}

\author[0000-0003-1051-6564]{Jenna M. Cann}
\altaffiliation{NASA Postdoctoral Program Fellow}
\affiliation{X-ray Astrophysics Laboratory, NASA Goddard Space Flight Center, Code 662, Greenbelt, MD 20771, USA}
\affiliation{Oak Ridge Associated Universities, NASA NPP Program, Oak Ridge, TN 37831, USA}

\author[0000-0002-4902-8077]{Nathan J. Secrest}
\affiliation{U.S. Naval Observatory, 3450 Massachusetts Avenue NW, Washington, DC 20392, USA}

\author[0000-0003-4701-8497]{Michael A. Reefe}
\altaffiliation{National Science Foundation, Graduate Research Fellow}
\affiliation{Department of Physics, Massachusetts Institute of Technology, Cambridge, MA 02139}
\affiliation{Kavli Institute for Astrophysics and Space Research, Massachusetts Institute of Technology, Cambridge, MA 02139}

\author[0000-0002-4375-254X]{Thomas Bohn}
\affiliation{Hiroshima Astrophysical Science Center, Hiroshima University, 1-3-1 Kagamiyama, Higashi-Hiroshima, Hiroshima 739-8526, Japan}



\begin{abstract}
Theoretical studies predict that the most significant growth of supermassive black holes occurs in late-stage mergers, coinciding with the manifestation of dual active galactic nuclei (AGNs), and both major and minor mergers are expected to be important for dual AGN growth. In fact, dual AGNs in minor mergers should be signposts for efficient minor merger-induced SMBH growth for both the more and less massive progenitor. We identified two candidate dual AGNs residing in apparent minor mergers with mass ratios of $\sim$1:7 and $\sim$1:30. SDSS fiber spectra show broad and narrow emission lines in the primary nuclei of each merger while only a narrow [\ion{O}{3}] emission line and a broad and prominent H$\alpha$/[\ion{N}{2}] complex is observed in the secondary nuclei. The FWHMs of the broad H$\alpha$ lines in the primary and secondary nuclei are inconsistent in each merger, suggesting that each nucleus in each merger hosts a Type~1 AGN. However, spatially-resolved LBT optical spectroscopy reveal rest-frame stellar absorption features, indicating the secondary sources are foreground stars and that the previously detected broad lines are likely the result of fiber spillover effects induced by the atmospheric seeing at the time of the SDSS observations. This study demonstrates for the first time that optical spectroscopic searches for Type~1/Type~1 pairs similarly suffer from fiber spillover effects as has been observed previously for Seyfert~2 dual AGN candidates. The presence of foreground stars may not have been clear if an instrument with more limited wavelength range or limited sensitivity had been used. 

\end{abstract}



\section{Introduction} 
\label{sec:intro}
Mergers of galaxies \citep[e.g.,][]{toomre1972,rothberg2004} are a ubiquitous phenomenon in the universe and represent a natural consequence of our current cosmological paradigm of ``bottom-up'' mass accumulation and galaxy evolution. During the merger process, tidal torques drive inflows of gas into the galaxy nuclei \citep[e.g.,][]{barnes1996} which can fuel both star formation \citep[e.g.,][]{barnes1991,mihos1996} and the growth of the central supermassive black holes \citep[SMBHs, e.g.,][]{hopkins2006,hopkins2008}, and this avenue for mass redistribution and transformation into stellar and/or SMBH mass has been postulated as one way the known scaling relations between the masses of the central SMBHs and the velocity dispersions \citep[M-$\sigma$, e.g.,][]{ferrarese2000,gebhardt2000} or the luminosities \citep[M-L, e.g.,][]{gultekin2009} of the host spheroids are established. It is generally accepted that the peak of active galactic nucleus (AGN) accretion activity -- and the point at which a merger should host a dual AGN -- should occur at nuclear separations of $<10$\,kpc based upon numerical simulations of major mergers \citep[mass ratios $<$1:3, e.g.,][]{vanwassenhove2012,capelo2015,blecha2018}, and simulations appear to show this is also the case for minor mergers (mass ratios $>$1:3) \citep{callegari2009,callegari2011,capelo2015}, the latter point of which is particularly interesting because minor mergers outnumber major mergers by a factor of $\geq3$--4 at late epochs \citep[e.g.,][and references therein]{kaviraj2014}.

Numerical and hydrodynamic simulations offer a relatively consistent physical picture of merger-induced SMBH growth and evolution in minor mergers, comprising three key features: (1) substantial growth of the SMBH in the less massive progenitor and the triggering of dual AGNs when the nuclear separations are $<10$\,kpc \citep{callegari2009,callegari2011,vanwassenhove2012,capelo2015}; (2) the drastic evolution of the merger mass ratio \citep{callegari2011,capelo2015}; and (3) the eventual formation of massive bound binaries \citep{callegari2009,callegari2011,khan2012b,capelo2015}. Although the average duty cycle for a dual AGN in a 1:10 merger is far shorter than in gas-rich major mergers \citep{callegari2011,vanwassenhove2012,capelo2015,capelo2017}, the SMBH hosted in the minor companion should grow 2- to 10-fold in mass during the evolution of the merger \citep{callegari2011,khan2012b,capelo2015}! This suggests that AGN growth in minor mergers could be an important SMBH growth channel and a pathway to synchronized AGN growth, and indeed a handful of dual AGNs have been observed in late-stage minor mergers: Was~49b \citep{bothun1989,moran1992,secrest2017}, J0924+0510 \citep{liu2018}, IRAS 05589+2828 \citep{koss2012}, and J1126+2944 \citep[a 460:1 mass ratio merger;][]{comerford2015}, all of which present as optical Type~2 dual AGNs.\footnote{Was~49b exhibits very broad H$\alpha$ \citep[$\sim6400$~km~s$^{-1}$;][]{secrest2017}, but this broad emission has a low equivalent width and is known to be highly polarized \citep[e.g.,][]{1995ApJ...440..578T}, indicating that it is likely scattered broad line emission.} However, there exists no confirmed case of a kiloparsec-scale dual AGN minor merger in which both nuclei harbor optically Type~1 AGNs.\footnote{\citet{foord2020} recently identified J2356-1016, a minor merger with at least one AGN \citep{pfeifle2019a}, as a dual AGN candidate based on Chandra X-ray imaging and SDSS optical spectra. Their analysis of the SDSS fiber spectra revealed tentative evidence for broad lines in the optical spectra of the both nuclei, though follow-up observations are required to confirm these results.} Dual Type~1/Type~1 AGNs have been observed at higher redshifts, but only in major (i.e. nearly equal mass) mergers \citep[e.g.,][]{silverman2020}. 

\begin{figure}[t!]
    \centering
    \subfloat{\includegraphics[width=.99\linewidth]{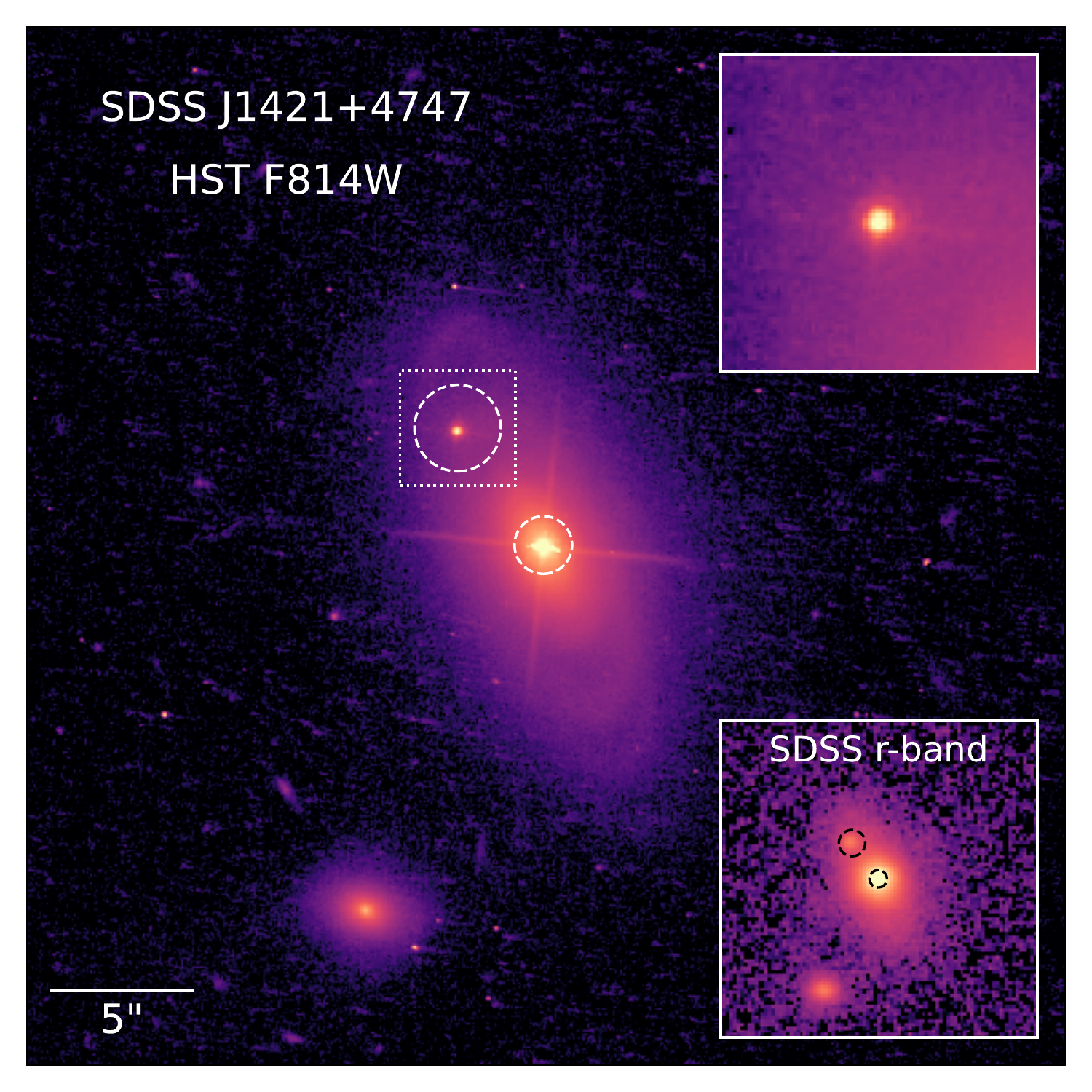}}\\
    \caption{The HST optical morphology of SDSS J1421+4747. (Left): The main figure displays the HST F814W image using a log scaling and a perceptually uniform color map. The scale bar in the lower left corner indicates a distance of 5$\arcsec$ or 6.9 kpc at $z=0.073$. The dashed white circles indicate the positions (and radii) of the SDSS fibers. The sources are separated by 7.2\,kpc ($5\farcs04$, based on the fiber positions) with a velocity offset of $|\Delta\,v|\approx180$\,km s$^{-1}$. The primary nucleus was observed with the BOSS spectrograph using a 2$\arcsec$ diameter fiber while the secondary nucleus was observed as a part of the SDSS Legacy Survey using a 3$\arcsec$ diameter fiber.  The lower right panel displays the lower resolution SDSS r-band image with the same scaling, color scheme, and zoom as the main panel, and the dashed black circles again indicate the SDSS fiber positions and sizes. The upper right panel shows a zoom-in view ($3\arcsec\times3\arcsec$) of the secondary source, indicated by the dotted square in the main panel.
    }
    \label{fig:j1421imaging}
\end{figure}

In a separate study of AGNs in mergers, we came across SDSS J142129.75+474724.5 (hereafter J1421+4747, $z=0.073$), which presents clear tidal features and consists of a primary galaxy with a bright central nucleus and an off-nuclear, dimmer secondary source (see Figure~\ref{fig:j1421imaging}), with a projected pair separation of $7.2$\,kpc ($5\arcsec$) and velocity offset of $|\Delta\,v|\approx 180$\,km s$^{-1}$. SDSS spectroscopic fiber measurements are available for both the primary nucleus and the offset, secondary source, revealing broad optical emission lines in both sources, highly suggestive of a dual Type~1/Type~1 AGN system. The bright, primary nucleus displays prominent, broad emission lines and a few narrow emission lines and is classified as a Seyfert~1.5 \citep[SBS 1419+480, e.g. ][]{baumgartner2013}. The dimmer, secondary source exhibits a strong host galaxy continuum, a narrow [\ion{O}{3}] emission line, and a very prominent and broad H$\alpha$/[\ion{N}{2}] emission line complex. Mass measurements from the SDSS and \cite{chen2018} indicate bulge masses of log$(M$/$M_\odot)\sim10.8$ and $\sim9.3$ for the primary and secondary sources, yielding a bulge mass ratio of $\sim1$:$29$, which---if true---would make J1421+4747 a dual Type~1/Type~1 AGN in a minor merger with one of the largest mass ratios of any known dual AGNs.

Although we stumbled upon J1421+4747 serendipitously, we used the SDSS specObj table for DR16 with a similar selection strategy to try and uncover further cases of Type~1/Type~1 dual AGNs in the local universe. We retrieved all groups of spectra within 1 arcmin and 1000 km s$^{-1}$ of each other that have ``BROAD'' in their SUBCLASS. We removed groups of spectra that corresponded to the same photo-objid (BESTOBJID), and then kept only those with projected separations less than 10 kpc. This left 60 pairs of spectra which were manually inspected. We removed those for which one or both members of the pair are classified as ``GALAXY BROADLINE'', since these are apparently absorption-dominated stellar spectra with no obvious broad emission lines. Finally, we were left with 13 systems, each with a pair of spectra that appear to both show broad emission features. This selection strategy recovered J1421+4747, the known dual AGN Was~49 \citep[e.g.,][]{secrest2017}, and the little-known dual AGN J1558+2723 \citep[the G3-G5 complex from][]{eckert2017}. One other system represented a convincing Type~1/Type~1 candidate, SDSS J171322.58+325627.9 (hereafter J1713+3256, $z=0.102$), which -- like J1421+4747 -- showed broad and narrow emission lines in the primary nucleus and a broad H$\alpha$/[\ion{N}{2}] complex and very narrow [\ion{O}{3}] narrow emission line in the secondary source; the SDSS measurements of the sources in J1713+3256 indicated a mass ratio of $\sim1$:$7$, suggesting it was also a dual Type~1/Type~1 AGN in a minor merger. The remaining 9 other systems selected from SDSS were not convincing Type~1/Type~1 candidates\footnote{See Appendix~\ref{appendixA} for a brief discussion of all 13 systems and why the remaining 9 targets were not viable targets for follow-up observations.}; four of these systems have been shown to suffer from fiber spillover effects \citep[e.g.,][]{husemann2020} and are likely single AGNs, while the remaining five systems have fiber positions that overlap and the resulting spectra are therefore likely to include substantial contributions from both nuclei.

However, given the $\lesssim5\arcsec$ separation of the nuclei in each system and the fact that the spectra were obtained using $3\arcsec$ or $2\arcsec$ diameter fibers, it is possible that the broad line observed in both secondary nuclei is spillover light from the central nuclei (as a result of inadequate atmospheric seeing), as has been shown to be an issue for narrow line emission in \citet[][]{husemann2020}. To investigate the true nature of these unique systems, we have reanalyzed the SDSS spectrum of each nucleus in each merger, and we have obtained and analyzed new, spatially resolved long slit optical spectra of J1421+4747 and J1713+3256 from the Large Binocular Telescope observatory (LBT) using the Multi Object Dispersion Spectrograph (MODS-2). We organize this work as follows: we describe the SDSS and LBT MODS-2 observations of each nucleus in Section~\ref{sec:observations} and we discuss our analysis of these spectra in Section~\ref{sec:analysis}. We present our spectroscopic results from the SDSS and LBT spectra in Section~\ref{sec:results}, discuss these results in the context of the literature in Section~\ref{sec:discussion}, and we provide our conclusions in Section~\ref{sec:conclusion}. We provide details on all 13 `Type~1/Type~1' systems selected from SDSS in Appendix~\ref{appendixA}. Throughout this work we assume the following cosmology: $\textrm{H}_0 = 70$\,km\,s$^{-1}$\,Mpc$^{-1}$, $\Omega_\mathrm{M}=0.3$, and $\Omega_\Lambda=0.7$; in this cosmology, $5\arcsec$ corresponds to a projected separation of 6.7~kpc at $z=0.07$.

\begin{figure}[t!]
    \centering
    \subfloat{\includegraphics[width=0.99\linewidth]{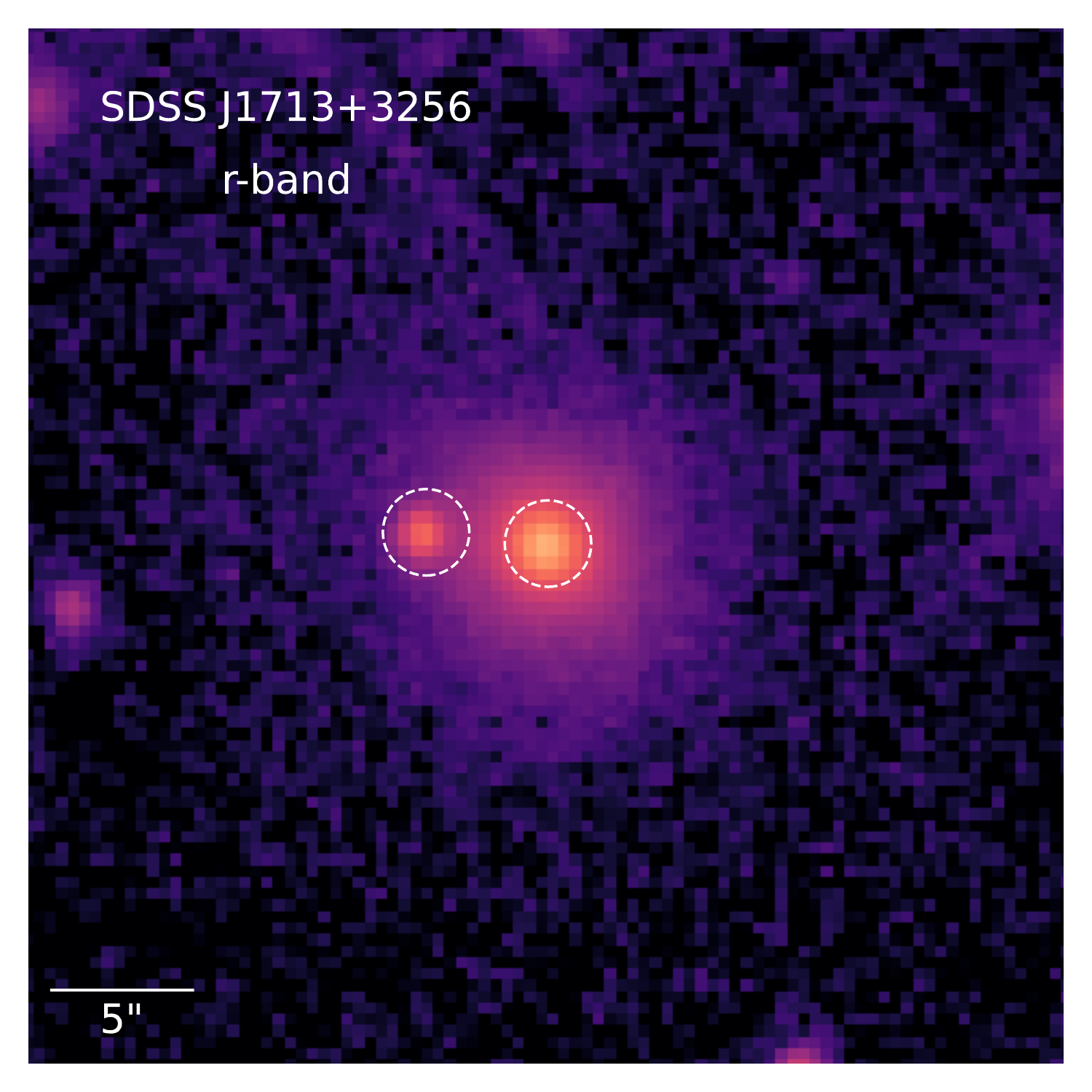}}
    \caption{SDSS Image and Optical Spectra for J1713+3256. (Left): The SDSS r-band image of J1713+3256 shows a larger, primary galaxy along with a smaller off-nuclear knot. The nuclei are separated by 7.9\,kpc ($4\farcs32$) with a velocity offset of $|\Delta\,v|\sim46$\,km s$^{-1}$. The scale bar indicates an angular size of 5$\arcsec$, or 9.2\,kpc at $z=0.1$. The dashed white circles indicate the positions (and radii) of SDSS fibers. Each nucleus was observed using a 3$\arcsec$ diameter fiber as part of the SDSS Legacy program.  
    }
    \label{fig:j1713imaging}
\end{figure}

\section{Observations and Data Processing}
\label{sec:observations}

\subsection{SDSS Spectroscopy}
\label{sec:sdssobs}
The primary nucleus in J1421+4747 was observed with the Baryon Oscillation Spectroscopic Survey (BOSS) spectrograph \citep{smee2013,dawson2013} on UT 17 March 2013 using a 2\arcsec{} diameter fiber (fiber ID 946, plate ID 6751) with median seeing $\sim1\farcs{}5$ (\textsc{SEEING50} entry from the `Plate' data table). The secondary nucleus in J1421+4747 was observed as a part of the SDSS Legacy Survey on UT 31 March 2005 using a 3\arcsec{} diameter fiber (fiber ID 55, plate ID 1672) with median seeing $\sim1\farcs{}8$. We show the fiber positions on the archival HST and SDSS imaging in Figure~\ref{fig:j1421imaging}.

Both the primary and secondary nuclei in J1713+3256 were observed as a part of the SDSS Legacy Survey. The primary nucleus was observed on 19 May 2002 UT using a 3\arcsec{} diameter fiber (fiber ID 619, plate ID, 976), while the secondary nucleus was observed on 05 May 2008 UT using a 3\arcsec{} diameter fiber (fiber ID 284, plate ID, 2973). Both the primary and secondary source in J1713+3256 were observed with a median seeing of $\sim1\farcs{}8$. We show the fiber positions on the SDSS imaging in Figure~\ref{fig:j1713imaging}.

\subsection{LBT Optical Spectroscopy}
\label{sec:lbtopticalobs}
Optical spectroscopy of J1421+4747 and J1713+3256 (Program ID: LD-2022A-003) were obtained with the Large Binocular Telescope (LBT) on Mt.Graham, Arizona, USA on MJD 59473 (UT 16 September 2021) and MJD 59671 (UT 02 April 2022).  LBT is a binocular telescope with two 8.4 meter primary mirrors residing on a single altitude-azimuth mount.  Each side contains a prime focus camera, four bent Gregorian f/15 ports, and direct f/15 Gregorian port. All observations used both optical Multi-Object Double Spectrographs, which are mounted on the direct Gregorian f/15 ports behind each primary mirror  \citep{2010SPIE.7735E..0AP}.  Each MODS contains a red and blue channel for spectroscopy, and both were used (Dual Grating mode) to provide coverage from  $0.32\,\mu\textnormal{m}-1.05\,\mu$m.  A 1$\arcsec$ wide longslit (R $\sim$ 900-1500) was used and each MODS were placed at a position angle so that both nuclei were aligned in the slit for each set of observations.  Two exposures of 1800 seconds were taken in each channel, however, the 2nd exposure for J1421+4747 was read out early due to the object reaching the elevation limits of the telescope.  The total integration times were 3100 seconds for J1421+4747 and 3600 seconds for J1713+3256 and the mean airmass of the observations were 1.99, and 1.02, respectively.  Observations of the spectro-photometric standard Feige 110 and BD+33 2642 were used to flux calibrate the data and remove the instrumental signatures from the data for J1421+4747 and J1713+3256, respectively.

The MODS data were reduced first with modsCCDRed \citep{2019zndo...2647501P} version 2.04 to remove bias and flat-field the data using a slitless pixel flat.  Next custom {\rm IRAF} scripts were used to:  rectify the tilt in both X and Y using a trace from the the spectrophotometric star and wavelength calibration from arc-lamp lines; correct the final wavelength calibration using known strong auroral skylines in the blue ([OI] $\lambda$=557.7338 nm) and red ([OI] $\lambda$=630.03 nm) channels; extract a one-dimensional spectrum from each channel; flux calibrate the data using the spectro-photometric standard star;  remove telluric features from the red channels using the normalized
spectro-photometric standard spectrum; and combine the red and blue channels of each MODS (including re-sampling both channels to a common value of 0.85 $\AA{}$ per pixel corresponding to an instrumental resolution of 5.8 $\AA{}$, and correcting the data to a heliocentric velocity).  Two different 1D apertures were extracted for each nucleus in each target. A 3$\arcsec$ diameter aperture was extracted, centered on the second "nucleus" for both targets.  This was to match and compare with the spectra from the SDSS fiber.  A 1$\arcsec$.25 diameter aperture was extracted for both ``nuclei'' in J1421+4747 and a 0$\arcsec$.65 diameter aperture was extracted for both ``nuclei'' in J1713+3256.  These aperture sizes were selected based on the mean seeing over the course of the observations as determined from telemetry from the off-axis guider and wavefront sensor for each MODS.  

After reviewing the final combined blue+red spectrum from each MODS, specifically  a visual comparison of the rest-frame stellar absorption lines in the second ``nucleus'' of each target,  it was decided to analyze only the data obtained with MODS-2 for both targets in order to maximize the S/N.  Combining the data from the two MODS resulted in reduced S/N in the final spectrum of the second object in the slit.  This is due to a known technical issue in which the sensitivity of MODS-1 has decreased by a factor of 1.6 since the 2011 commissioning\footnote{https://scienceops.lbto.org/mods/preparing-to-observe/sensitivity/}.\footnote{The cause of this degradation has yet to be localized to a single source, and may be some combination of reduced sensitivity of various optical components in MODS-1 and reduced reflectivity at optical wavelengths from the Adaptive Secondary Mirror coating.}

\section{Data Analysis}
\label{sec:analysis}
To fit the spectra from SDSS and LBT, we used the open-source spectral analysis program Bayesian AGN Decomposition Analysis for SDSS Spectra \citep[BADASS;][]{sexton2021}.\footnote{https://github.com/remingtonsexton/BADASS3} BADASS uses the affine-invariant Markov-Chain Monte Carlo sampler \textsc{emcee} for robust parameter and uncertainty estimation, and it can fit simultaneously for a variety of spectral features, including: individual narrow, broad, and/or absorption line features, the stellar line-of-sight velocity distribution (LOSVD) through the use the penalized PiXel-Fitting (pPXF) method from \citet[][]{cappellari2017}, broad and narrow \ion{Fe}{2} emission features, the AGN power law continuum, and blue-wing outflow emission components.

For each nucleus we model the narrow and broad emission line profiles using simple Gaussians, and we use the default settings of BADASS when possible \citep[see Section~2.1.2 in][for a brief description of the default line parameter settings, including line parameters that are tied across different emission lines]{sexton2021}. For the primary nucleus in J1421+4747, we also modeled additional blue-shifted components as simple Gaussians; we refer to these components as outflows throughout the remainder of this work, but it is important to note that the complex shape of the lines could be due to outflows in the narrow components, complex kinematics in the broad line component, the specific viewing angles, or a combination of the these effects. For primary nuclei in J1421+4747 and J1713+3256, the S/N of the spectra allowed us to fit for the stellar LOSVD and the AGN continua. Given the lower S/N of the continuum in each of the secondary nuclei, we elected to fit only for the broad and narrow emission lines, the intrinsic AGN power law continuum, and the broad and narrow \ion{Fe}{2} optical lines and we did not fit for the stellar LOSVD or for the host stellar population template. In a few cases, such as the H$\beta$ emission line in the spectrum of the primary nucleus of J1421+4747, simple Gaussians do not well reproduce the shape of the broad, narrow, and outflow components of the line; BADASS does allow the user to choose other line profiles---such as Voigt or Gauss-Hermite profiles---but as we show in Section~\ref{sec:results}, simple Gaussian profiles are sufficient for this study. To reduce computation times, we fit the [\ion{O}{3}] and H$\alpha$/[\ion{N}{2}] regions separately (4400-5500\AA{} and 6200-6900\AA{}, respectively).

\section{Results}
\label{sec:results}
In fitting the SDSS spectra of these candidate dual AGNs, we were chiefly interested in confirming the presence of broad H$\alpha$ emission lines within the secondary nuclei, and we therefore measure the full-width at half maximum (FWHM) of the broad lines to compare the kinematics of the lines in the primary and secondary nuclei. In the case of J1421+4747, we used the line testing feature in BADASS to find that a broad H$\alpha$ line was needed to reproduce the observed line with $>99.9$\% confidence based on the BADASS A/B likelihood test (significance $>3\sigma$). We found FWHMs of the broad H$\alpha$ emission lines of $6073^{+24}_{-46}$\,km~s$^{-1}$ and $4290^{+119}_{-118}$\,km~s$^{-1}$ in the primary and secondary nuclei in J1421+4747, where these FWHM values differ at the 3$\sigma$ level\footnote{However, these uncertainties are derived from posteriors corresponding to the assumed Gaussian line model, so these uncertainties may be underestimated; see Section~3.2 from \citet{secrest2017}.}, suggesting physically distinct kinematic broad line regions, and thus suggest physically distinct emission regions (i.e. two distinct BLRs). For J1713+3256, we found that the broad emission line in the secondary source was detected with only $\sim48$\% confidence and a significance of only $\sim0.5\sigma$ based on the A/B likelihood test (suggesting a misclassification by the SDSS pipeline, but the S/N of the spectrum warranted follow-up observations); the FWHM values of the broad H$\alpha$ emission lines in the primary ($4649^{+55}_{-84}$\,km~s$^{-1}$) and secondary ($3560^{+318}_{-190}$\,km~s$^{-1}$) nuclei differ at the $\sim2.7\sigma$ level, suggesting that J1713+3256 also hosts two kinematically distinct broad line regions (if the broad line in the secondary nucleus is real). The FWHM values derived for these two systems are tabulated in Table~\ref{table:SDSSresults}; given the inconsistent widths of the broad H$\alpha$ emission lines between the primary and secondary nuclei within each system, the SDSS spectra would suggest we are observing distinct broad line regions associated with each source in each merger, providing evidence for the presence of dual Type~1 AGNs. 

\begin{figure}
    \centering
    \subfloat{\includegraphics[width=0.99\linewidth]{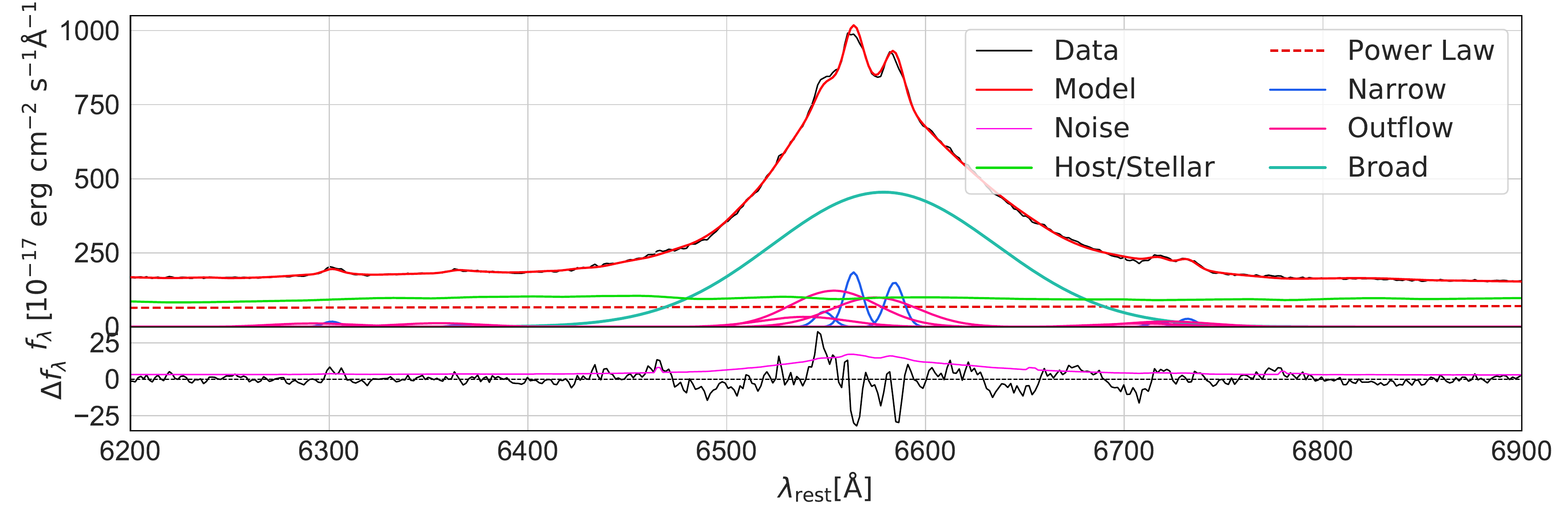}}\\ \vspace{-0.2cm}
    \subfloat{\includegraphics[width=0.99\linewidth]{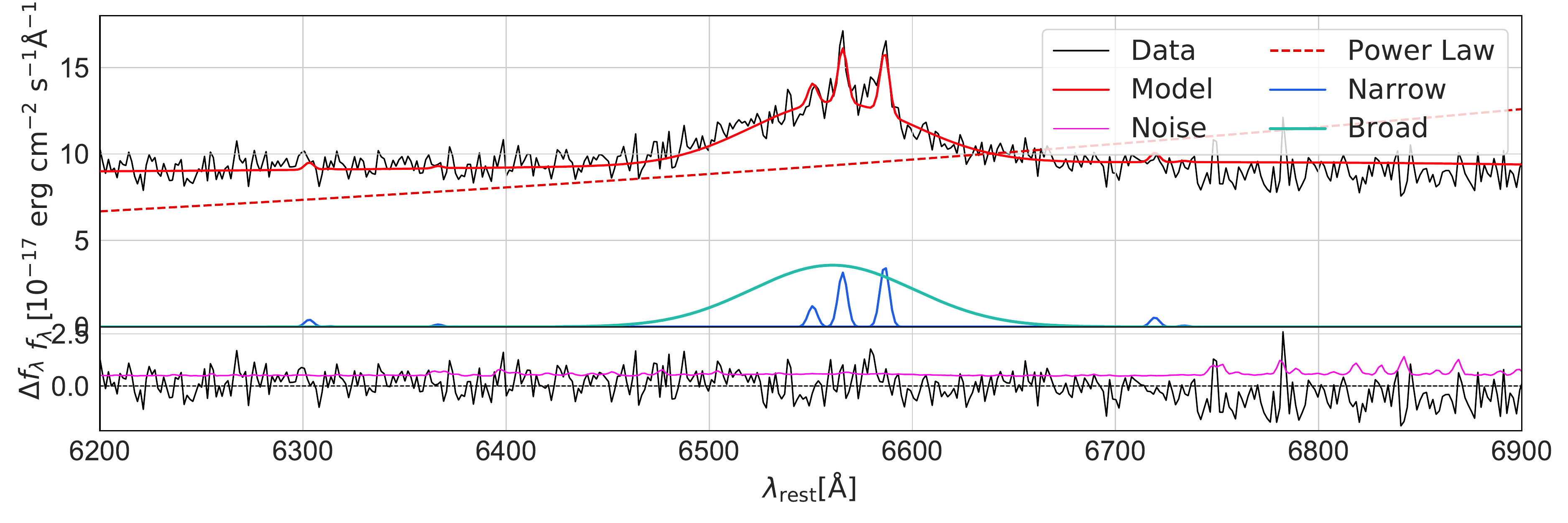}}\\\vspace{-0.2cm}
    \subfloat{\includegraphics[width=0.99\linewidth]{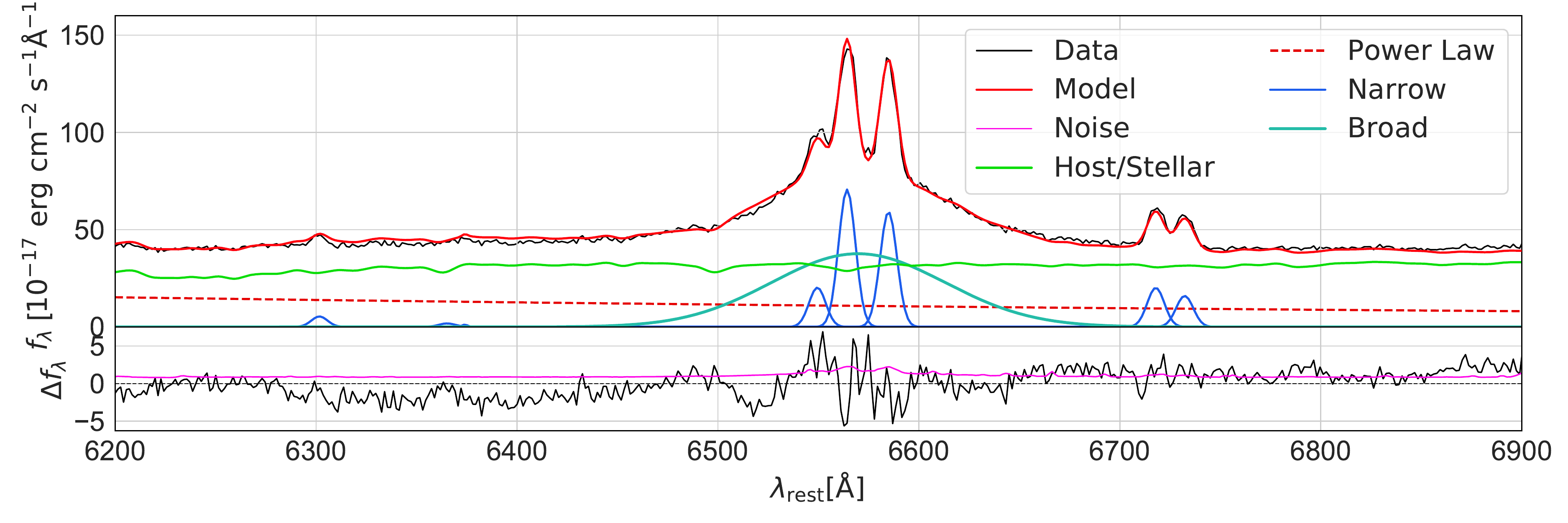}}\\\vspace{-0.2cm}
    \subfloat{\includegraphics[width=0.99\linewidth]{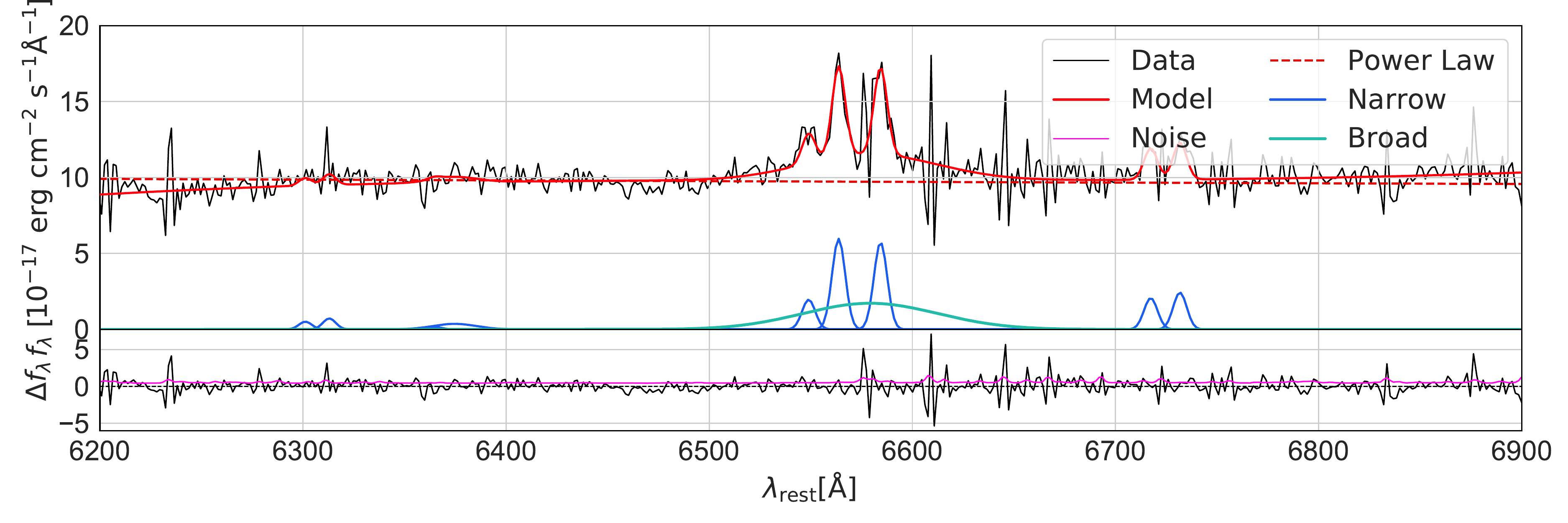}}\\
    \caption{Rest-frame SDSS fiber spectra fit with BADASS. Top to bottom: the H$\alpha$$\lambda$6563/[N II]$\lambda$6584 spectral region for J1421+4747 primary, J1421+4747 secondary, J1713+3256 primary, J1713+3256 secondary. The various spectral components are listed in the legends in the top right corner of each plot; the final total spectral fit is given as the red line tracing the SDSS spectrum. The [S II]$\lambda\lambda$6717,6733 doublet is also visible in the primary sources in J1421+4747 and J1713+3256.} 
    \label{fig:sdssfits}
\end{figure}

Going beyond the broad H$\alpha$ emission line observed in the SDSS spectrum of each nucleus, we also examined the optical spectroscopic emission line ratios \citep{baldwin1981,kewley2001,kauffmann2003,kewley2006} derived from each spectrum; we quote only upper or lower limits for the [\ion{O}{3}]/H$\beta$, [\ion{S}{2}]/H$\alpha$, and [\ion{O}{1}]/H$\alpha$ emission line ratios in the secondary nuclei due to the lack of detections (or strong detections) of the H$\beta$, [\ion{S}{2}], and [\ion{O}{1}] lines in the secondary spectra. In each case, the observed line ratios (including the upper/lower limits) place the nuclei within the Seyfert region of each BPT diagram, suggesting once again the presence of multiple AGN in each system. We also searched the spectra of the secondary sources for evidence of rest-frame stellar absorption features to rule out a scenario in which each pair comprises a foreground star and background AGN; we identified potential rest-frame Na D absorption lines (rest-frame $\lambda$5890 and $\lambda$5896) in both of the secondary sources, but given the S/N of the spectra and the lack of additional absorption features seen in the SDSS spectra, it was unclear if these absorption lines were genuine. Follow-up observations were needed to elucidate this issue, and for the time being we disfavored the foreground star-background AGN scenario.

\begin{figure}
    \centering
    \subfloat{\includegraphics[width=1.0\linewidth]{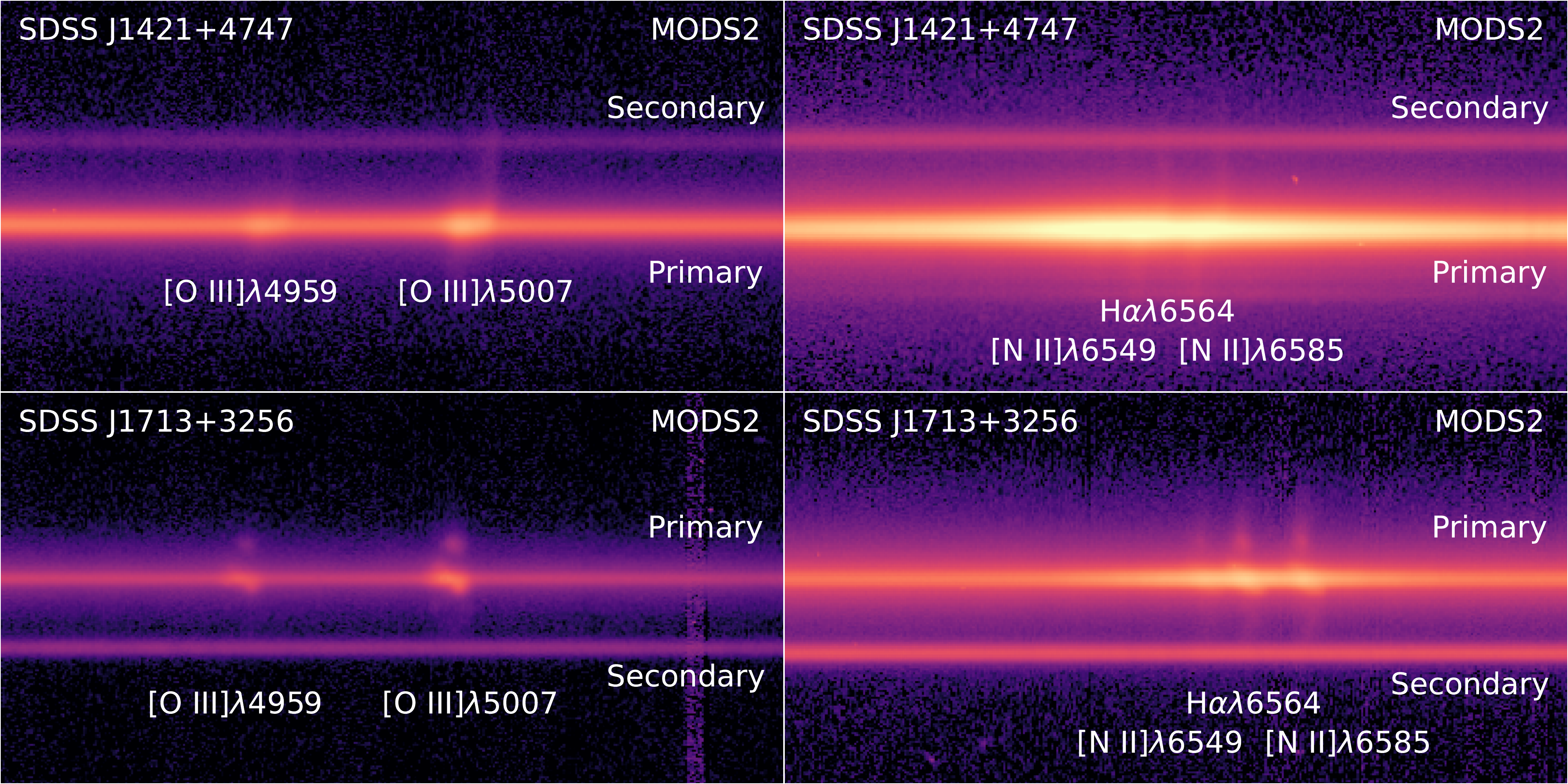}}\\
    \caption{LBT 2D optical spectra for (top) J1421+4747 and (bottom) J1713+3256. (Left) The spectral region around the [\ion{O}{3}]$\lambda$5007,$\lambda$4959 doublet. (Right) The spectral region around the H$\alpha$/[\ion{N}{2}] emission line region. Strong emission lines are observed in the spectra of the primary nuclei, while fainter and fairly featureless continua are observed in the secondary sources. Extended emission is observed and originates from the primary nuclei.}
    \label{fig:lbt2d}
\end{figure}

\begin{figure}
    \centering
    \subfloat{\includegraphics[width=1.0\linewidth]{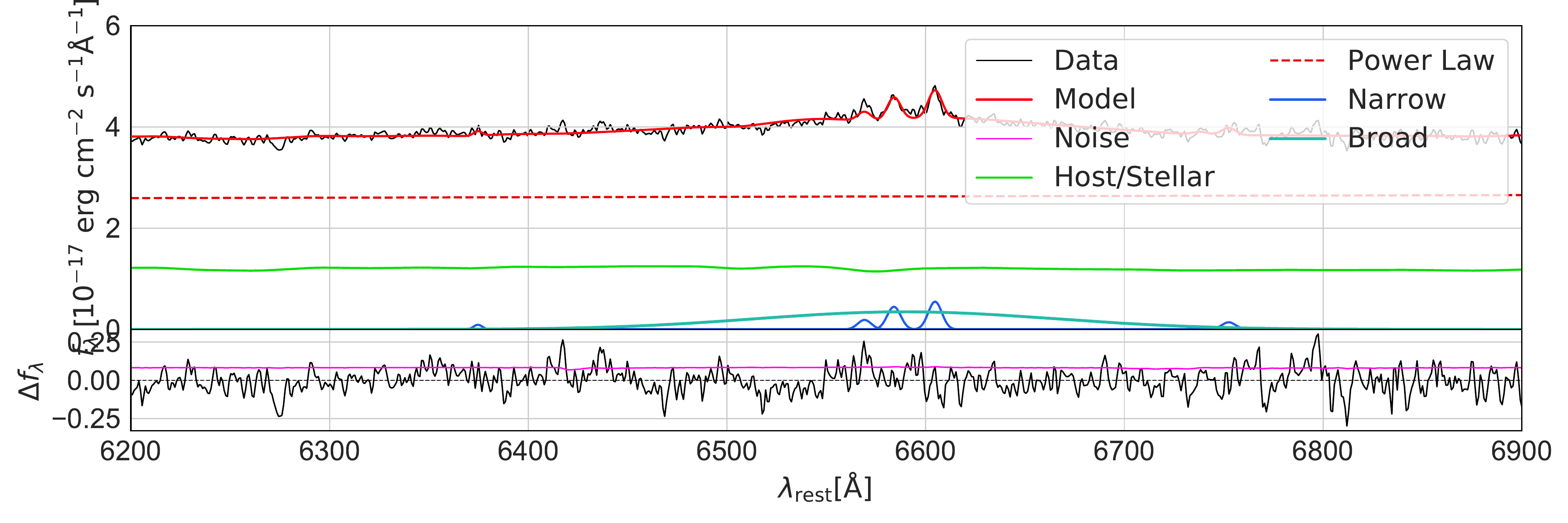}}\\
    \subfloat{\includegraphics[width=1.0\linewidth]{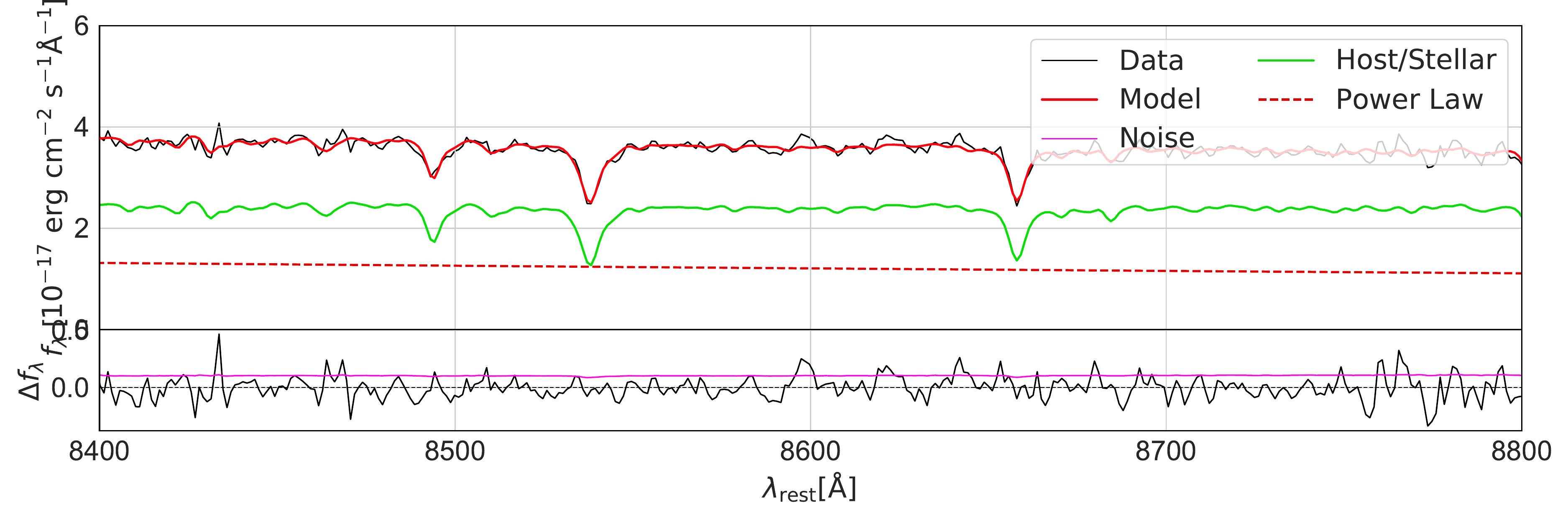}}\\
    \subfloat{\includegraphics[width=1.0\linewidth]{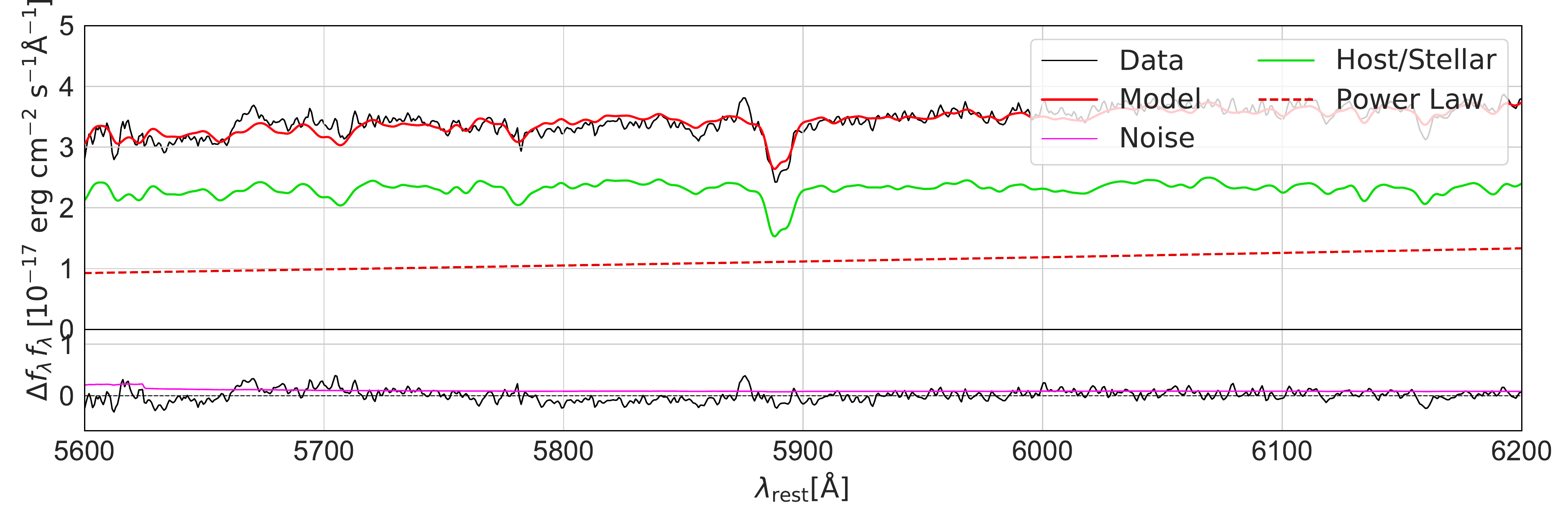}}\\
    \caption{LBT optical 1D spectra of the secondary source in J1421+4747. Top: the H$\alpha$ spectral region for J1421+4747 secondary, which shows the BADASS fit to the continuum and observed emission lines, including a broad H$\alpha$ line and a narrow H$\alpha$/[\ion{N}{2}] emission line complex. Middle and bottom: the detected rest-frame Ca triplet and Na D absorption line systems, indicating that the secondary source is in fact a foreground star and the observed emission lines in the H$\alpha$ region in both the SDSS and LBT data arise from spillover from the primary nucleus.
    }
    \label{fig:lbtspectra_j1421}
\end{figure}

\begin{figure}
    \centering
    \subfloat{\includegraphics[width=1.0\linewidth]{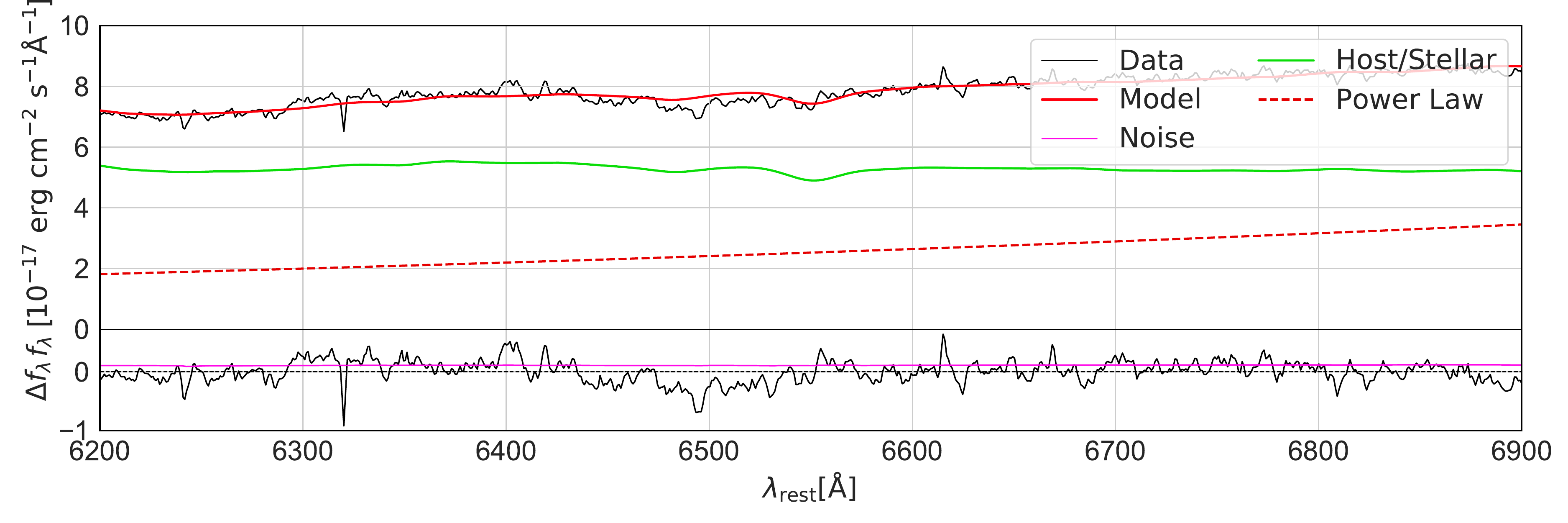}}\\
    \subfloat{\includegraphics[width=1.0\linewidth]{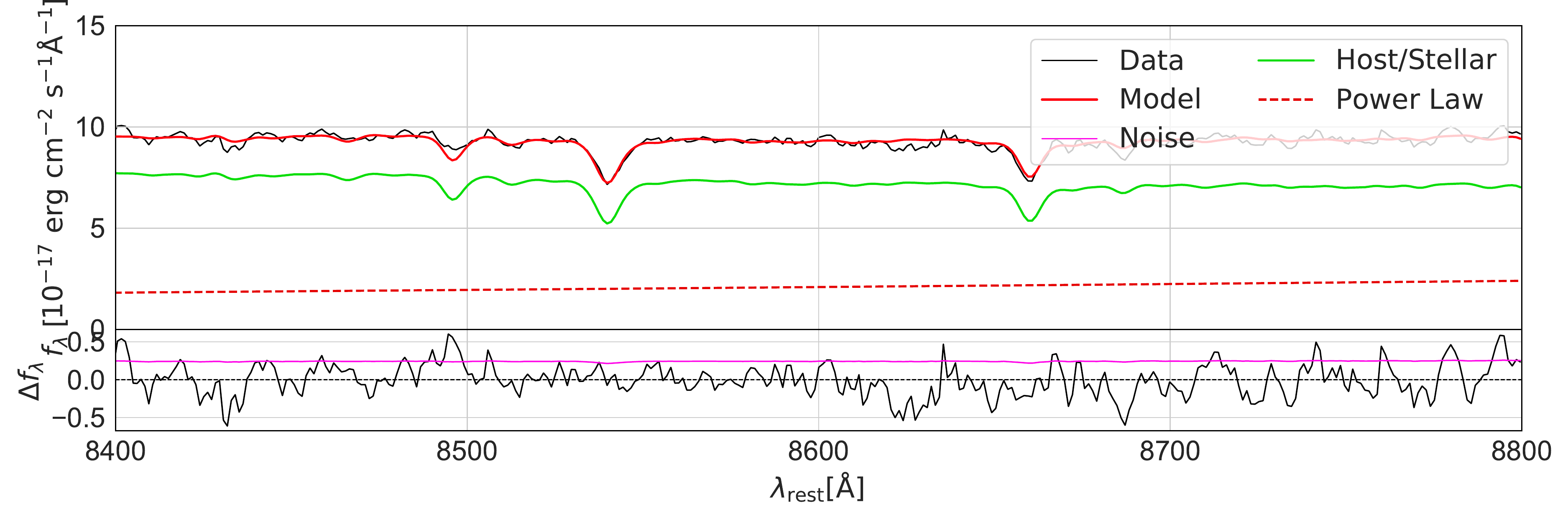}}\\
    \subfloat{\includegraphics[width=1.0\linewidth]{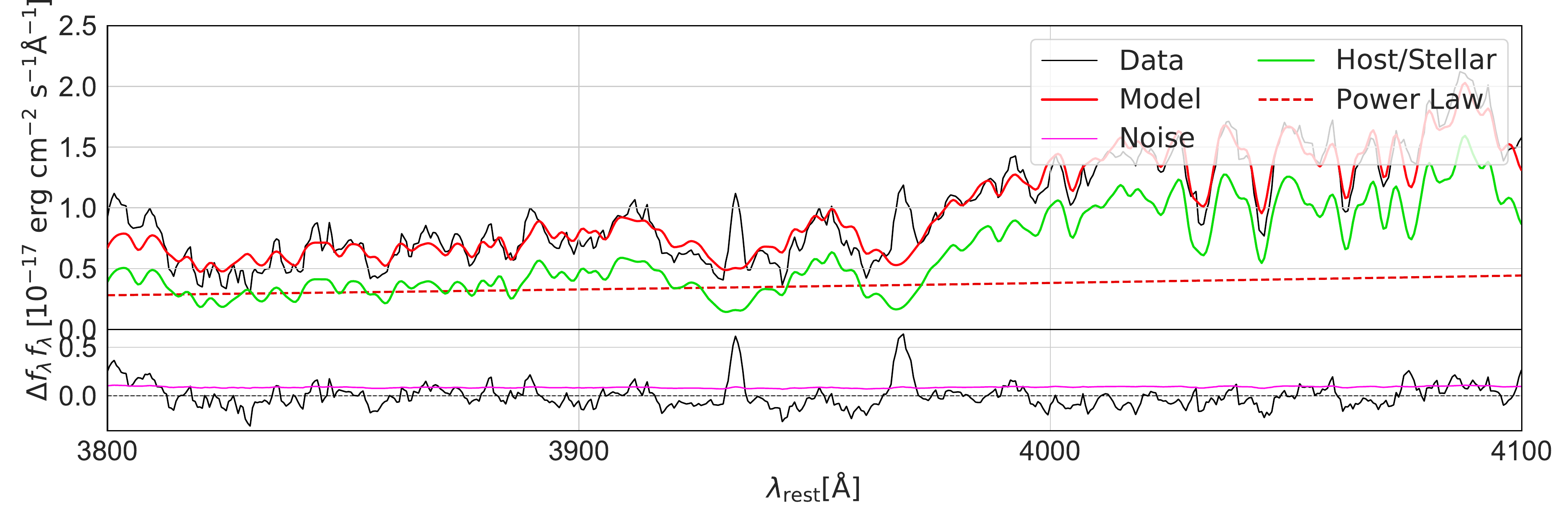}}\\
    \subfloat{\includegraphics[width=1.0\linewidth]{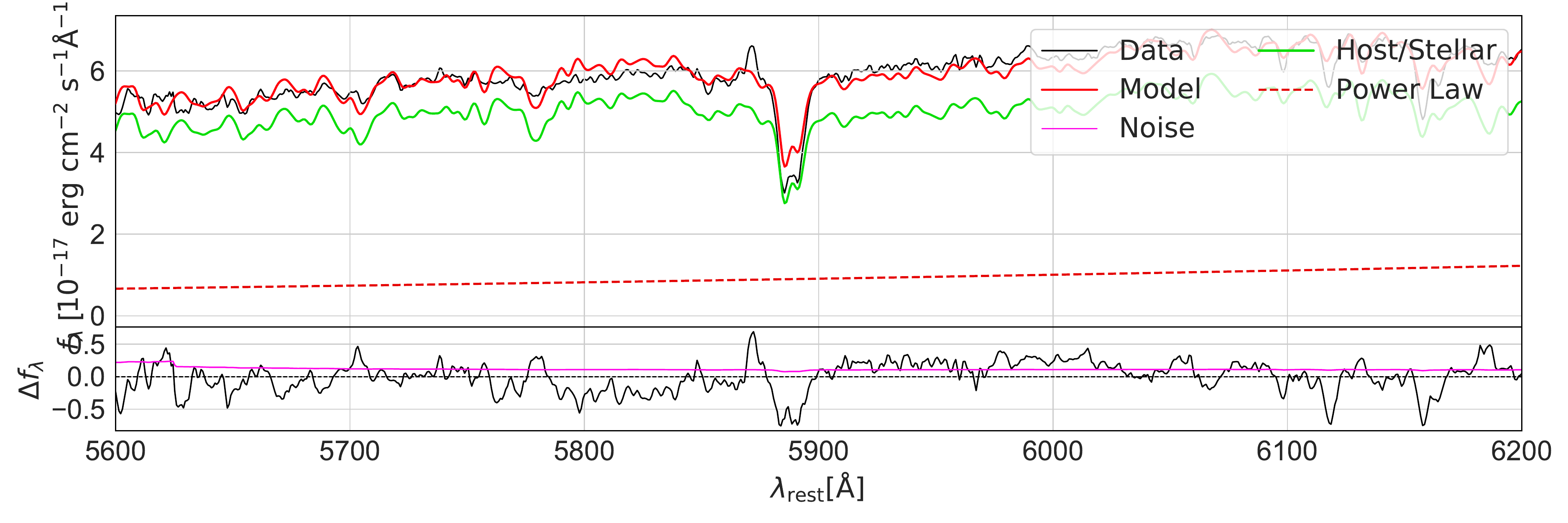}}\\
    \caption{LBT optical 1D spectra of the secondary source in J1713+3256. Top: the H$\alpha$ spectral region for J1713+3256 secondary, which shows the BADASS fit to a relatively featureless continuum. Top-middle: the detected rest-frame Ca triplet absorption line system along with the BADASS spectral fit. Bottom-middle: the detected rest-frame Ca K+H absorption lines along with the BADASS spectral fit. Bottom: the detected rest-frame Na D absorption lines along with the BADASS spectral fit. The middle two panels and the bottom panel indicate that the secondary source is a foreground star and not an AGN, unlike the conclusions made based on the SDSS spectra alone.}
    \label{fig:lbtspectra_j1713}
\end{figure}

At first glance, these results would suggest evidence for true Type~1 dual AGNs, but the true nature of these systems was only unveiled through the use of higher spatial resolution LBT spectroscopic observations.\footnote{The spectral resolution of our LBT spectra was lower than that of the SDSS spectra due to our use of a 1\arcsec{} slit, but this does not affect our results. Spatial, rather than spectral, resolution was the critical requirement for this work.} We show the 2D spectra for J1421+4747 and J1713+3256 in Figure~\ref{fig:lbt2d} around the H$\alpha$/[\ion{N}{3}] and [\ion{O}{3}] emission line regions; strong emission lines in the primary nuclei are present, while the secondary sources show only a strong continuum without obvious signs of emission lines. We extracted the 1D spectra of the secondary sources using a $1\farcs25$ diameter aperture in the case of J1421+4747 (Figure~\ref{fig:lbtspectra_j1421}) and a $0\farcs65$ diameter aperture in the case of J1713+3256 (Figure~\ref{fig:lbtspectra_j1713}) and we used BADASS to fit the 1D spectra and test the significance of broad and narrow line components. In J1421+4747, we again identified a broad H$\alpha$ line ($99.0$\% confidence based on the A/B likelihood test), although the H$\alpha$/[\ion{N}{2}] emission line complex is offset by roughly 30\AA{} redward of what was observed in the SDSS spectrum. More puzzling is the fact that the FWHM of the broad H$\alpha$ line was found to be $8064_{-443}^{+445}$\,km\,s$^{-1}$, nearly twice that found when analyzing the SDSS spectrum. 

While the broad H$\alpha$ detection on its own may have offered strong evidence that this system was a true dual AGN (but ignoring the unusually large discrepancy between the SDSS and LBT FWHM values), the 1D spectrum also revealed a rest-frame Ca triplet absorption line system (rest-frame 8500\AA{}, 8544\AA{}, and 8664\AA{})\footnote{These lines are offset blueward by $\sim6\AA{}$ from their rest-frame wavelengths, but this offset may not be significant given the instrumental resolution of 5.8\AA{}}, indicating that the continuum most likely arises from a foreground star and that the broad H$\alpha$ line and H$\alpha$/[\ion{N}{2}] complex are likely spillover from the primary nucleus. In the case of J1713+3256, the 1D spectrum lacks any characteristic signatures of Type~1 or Type~2 AGNs, and instead displays characteristic rest-frame stellar absorption features: a rest-frame Ca triplet absorption line system and rest-frame Ca K+H absorption lines (rest-frame 3934\AA{} and 3969\AA{}). Furthermore, we confirmed the presence of rest-frame Na D absorption lines (rest-frame $\lambda$5890 and $\lambda$5896) in the 1D spectra of the secondary sources in both J1421+4747 and J1713+3256. Thus, the LBT data offer strong evidence that the secondary sources are in fact foreground stars and that each of these systems unfortunately represent the chance alignment of a foreground star and background AGN that only presented as a dual AGN candidate as a result of inadequate atmospheric seeing at the time of the original observations. To confirm this, we matched the secondary sources onto Gaia~DR3 \citep{2022arXiv220800211G}, finding that both exhibit highly significant proper motions of 11.8~mas~yr$^{-1}$ for J1421+4747 and $10.3$~mas~yr$^{-1}$ for J1713+3256. While apparent proper motions have recently been linked to AGN multiplicity as well as source extent in low-$z$ systems \citep{2022A&A...660A..16S, 2022ApJ...933...28M}, the spectroscopic and astrometric evidence taken together strongly favors the secondary sources being foreground stars.

But what of the previously identified broad line in the secondary source of J1713+3256? The LBT 2D spectra gave us an initial clue: there is clear extended emission originating from the primary sources in both J1421+4747 and J1713+3256 that overlaps the continua of the secondary sources; it is this very extended emission that likely gives rise to the H$\alpha$/[\ion{N}{2}] complex still observed in the secondary source in J1421+4747, while the observation of J1713+3256 was taken under better seeing conditions and does not suffer as drastically. In an attempt to reproduce what is seen in the SDSS fiber spectra of J1713+3256, we re-extracted the 1D spectra of the secondary sources using a 3\arcsec{} diameter aperture (matched in size to the SDSS fibers) and refit the spectrum using BADASS. The broad H$\alpha$ and narrow H$\alpha$/[\ion{N}{2}] emission lines still remain undetected; this is likely due to the fact that the original SDSS spectra were taken in far worse observing conditions (median seeing of $\sim1\farcs{}8$ for both the primary and secondary nuclei of J1713+3256), while the LBT data were taken in nearly pristine conditions (average seeing $0\farcs{}5\pm0\farcs{}03$ for J1713+3256).

A possible explanation for the discrepant broad H$\alpha$ FWHM values – when comparing the two nuclei in SDSS (as well as when comparing the SDSS FWHM to that measured with LBT in the case of J1421+4747) -- could lie in (1) the size of the extraction apertures and (2) the complex spectral shape and the spatial distribution of the H$\alpha$ emission (as seen in the 2D spectra): In the case of SDSS, the poorer seeing conditions led to the smearing of the primary nuclear emission, allowing it to be observed in the secondary fiber, but the 3’’ diameter fibers are also effectively averaging emission from a large portion of the galaxy (offset from the nucleus), complicating the measured FWHM and resulting in values that were discrepant with the primary source. With LBT, we are likely better resolving the complex emission from several kinematically distinct spots within the background galaxy but with less contamination from the primary nucleus; however, the LBT measurement of the secondary source still comprises a combination of the foreground star and spatially resolved emission from areas outside of the galaxy's nucleus. It may not be surprising then, that the SDSS and LBT broad H$\alpha$ FWHMs do not match.

As for the optical classifications based upon the BPT narrow emission line ratios, these AGN signatures are undoubtedly the result of the fiber spillover contamination \citep[e.g.,][]{husemann2020}. However, other phenomenon such as extended narrow line regions \citep[e.g.,][]{haineline2014} or cross-ionization \citep[specifically when two galaxies are involved, e.g.,][]{keel2019} could also mimic dual AGN signatures.

\begin{table*}[ht]
\begin{center}
\caption{SDSS Fitting Results}
\label{table:SDSSresults}
\begin{tabular}{ccccccc}
\hline
\hline
\noalign{\smallskip}
\noalign{\smallskip}
Target & Nucleus & H$\alpha_{\rm{BR, FWHM}}$ & log($\frac{\rm{[OIII]}}{\rm{H}\beta}$) & log($\frac{\rm{[NII]}}{\rm{H}\alpha}$) & log($\frac{\rm{[SII]}}{\rm{H}\alpha}$) & log($\frac{\rm{[OI]}}{\rm{H}\alpha}$)  \\
(1) & (2) & (3) & (4) & (5) & (6) & (7)\\
\noalign{\smallskip}
\noalign{\smallskip}
\hline
\noalign{\smallskip}
J1421+4747 & 1 & $6073^{+24}_{-46}$ km~s$^{-1}$ &   $0.72^{+0.02}_{-0.03}$  & $-0.1^{+0.03}_{-0.04}$ & $-0.61^{+0.03}_{-0.05}$ & $-0.9^{+0.03}_{-0.05}$ \\
J1421+4747 & 2 & $4290^{+119}_{-118}$ km~s$^{-1}$ & $<0.73$  & $0.05^{+0.14}_{-0.09}$ & $>-0.56$ & $>-0.55$ \\ 
J1713+3256 & 1 & $4649^{+55}_{-84}$ km~s$^{-1}$ & $0.9^{+0.04}_{-0.03}$ & $-0.07^{+0.02}_{-0.01}$  & $-0.28^{+0.02}_{-0.02}$ &  $-1.03^{+0.05}_{-0.04}$ \\
J1713+3256 & 2 & $3560^{+318}_{-190}$ km~s$^{-1}$ & $<0.99$ & $0.04^{+0.1}_{-0.09}$  & $>-0.05$ & $>-0.77$  \\ 

\noalign{\smallskip}
\hline
\end{tabular}
\end{center}
\tablecomments{Col 1-2: merger designation and nucleus designation. Col 2: FWHM of the broad H$\alpha$ emission line, in units of km~s$^{-1}$. Col 3-7: [\ion{O}{3}]$\lambda$5007/H$\beta$, [\ion{N}{2}]/H$\alpha$, [\ion{S}{2}]$\lambda\lambda$6717,6733/H$\alpha$ doublet, and [\ion{O}{1}]$\lambda$6302/H$\alpha$ emission line ratios derived from the SDSS spectra using BADASS. Upper limits are given for the [\ion{O}{3}]$\lambda$5007/H$\beta$ line ratios of the secondary sources since H$\beta$ was not observed in these cases. Lower limits are given for the [\ion{S}{2}]$\lambda\lambda$6717,6733/H$\alpha$ doublet, and [\ion{O}{1}]$\lambda$6302/H$\alpha$ emission line ratios in the secondary sources due to a lack of strong [\ion{S}{2}] or [\ion{O}{1}] emission lines. All error bounds are quoted at 1$\sigma$. }
\end{table*}

\section{Discussion}
\label{sec:discussion}
\subsection{The Necessity of Spatially Resolved Spectroscopy}
Though an analysis of the SDSS optical fiber spectra alone would suggest the presence of dual Type~1/Type~1 AGNs in these presumed minor mergers, our follow-up LBT MODS-2 spectroscopic observations have now clearly shown that these systems comprise foreground stars and background AGNs caught in projection. A foreground star scenario was originally disfavored in both cases due to the absence of unambiguous rest-frame stellar absorption features (candidate Na D absorption lines were spotted in both secondary sources, but follow-up observations were needed to investigate this); a background AGN scenario was ruled out by the lack of lines significantly redshifted relative to the primary galaxy in each system. The measured FWHM values of the broad H$\alpha$ components were inconsistent when comparing the primary nuclei to the secondary nuclei, but the SDSS optical spectra alone could not definitively exclude the possibility that the fiber spectra of the secondary nuclei were contaminated by light from the primary nuclei; spatially resolved spectroscopy was required to falsify the original result. This work has clearly demonstrated -- for the first time for dual AGN candidates -- that fiber spillover for broad line AGNs can be a significant contaminant in spectroscopic campaigns searching for closely separated dual AGNs, as suggested by \citet{husemann2020} in a study on fiber spillover in narrow line dual AGN candidates. Furthermore, the results of this study emphasizes the need for optimal atmospheric seeing conditions when obtaining spectroscopic observations of closely separated sources in order to avoid significant fiber spillover.

\citet{husemann2020} developed a simple model using a 2D Moffat function to quantify the effect of fiber spillover via the flux ratio of two sources ($f_{primary}/f_{secondary}$) as a function of the radius between the two sources, the seeing, and $\beta$ parameter \citep[Figure~2 in][]{husemann2020}. This toy model offers an excellent check against fiber spillover for broad line AGNs as examined here: the SDSS flux ratios for the narrow [\ion{O}{3}] fluxes and the broad H$\alpha$ fluxes in J1421+4747 were
log10($F_{\rm{H\alpha,P}}/F_{\rm{H\alpha,S}}$) = 2.253 and 
log10($F_{\rm{[OIII],P}}/F_{\rm{[OIII],S}}$) = 2.131, and for J1713+3256 we found
log10($F_{\rm{H\alpha,P}}/F_{\rm{H\alpha,S}}$) = 1.421 and log10($F_{\rm{[OIII],P}}/F_{\rm{[OIII],S}}$) = 1.092; taking into account the fiber separations ($\sim4$--$5\arcsec$) in these systems, J1421+4747 and J1713+3256 both occupy the exact parameter space where fiber spillover effects are expected. These observations serve as a stark reminder that selection techniques relying upon spectroscopic fiber measurements must be carefully scrutinized in order to avoid source confusion and false positives. Taken together with our discussion in Appendix~\ref{appendixA} of the other `Type 1/Type 1' candidates selected from SDSS (most of which are, or are likely to be, the result of fiber spillover), this work emphasizes that fiber spillover of single AGN emission in poorer atmospheric seeing conditions is more commonplace than dual Type 1/Type 1 AGN systems. It should be noted that we obtained our optical spectra during higher quality observing conditions than those in \citet{husemann2020}, who also used LBT MODS spectroscopy with 1\arcsec{} wide slits \citep[see Section~\ref{sec:lbtopticalobs} above, as well as Table~1 in][for a comparison of the seeing conditions between their work and this work]{husemann2020}; our work confirms the effectiveness of their toy model for fiber spillover.

\subsection{Dual AGNs in Minor Mergers: A Perspective}
For simulated prograde-prograde, gas rich minor mergers with mass ratios of 1:4, 1:6, and 1:10, the growth of the SMBH in the minor companion results from approximately two phases: (1) tidally induced gas inflows---as a result of interactions with the more massive galaxy---lead to the build up of SMBH mass and stellar mass (via star formation) in the core of the less massive galaxy until ram pressure stripping abruptly halts this growth phase, and (2) the rapid fueling of the secondary SMBH once it and its associated stellar core have circularized within the disk of the primary and begin sweeping up vast reservoirs of gas \citep{callegari2009,callegari2011,vanwassenhove2012,capelo2015}. While the fueling of the more massive SMBH is expected to be stochastic during earlier pericenter passages, its growth pathway transitions from secular to merger-induced as the orbit of the secondary galaxy shrinks and it circularizes within the disk of the massive companion  \citep{callegari2009,callegari2011,capelo2015}. Thus, it is in the late-stage merger phase where we should expect to find correlated growth between the two SMBHs \citep{vanwassenhove2012}, and indeed most dual AGNs in minor mergers to date have been found in late-stage mergers \citep[][]{koss2012,comerford2015,secrest2017,liu2018} with the exception of Mrk 268 and NGC 1052/NGC 1042 \citep[where the nuclei are separated by $\sim$44\,kpc and $\sim$84\,kpc,][]{koss2012}. 

The SMBH mass ratios in these minor mergers can experience a dramatic evolution across the merger sequence: by the time the SMBHs have formed a bound binary, an initial 1:10 mass ratio has been shown to evolve to 1:6, while 1:6 and 1:4 mass ratios have been shown to evolve to 1:2 or larger \citep{callegari2011,capelo2015}. Astoundingly, 1:10 mass ratio mergers can even briefly evolve to 1:3 mergers during the merger phase \citep[but prior to the bound binary phase][]{callegari2011,capelo2015}; these striking results demonstrate not only that minor mergers can be incredibly efficient avenues for the build up and evolution of the secondary SMBH but that they also effectively erase the initial SMBH mass ratio. This latter point would suggest that some fraction of SMBH pairs with mass ratios in the range 1:1-1:3 may not have originated from major mergers at all, but rather could have arisen from minor mergers with significantly smaller initial SMBH mass ratios; the same point can be made for dual AGNs in minor mergers in which the SMBH mass ratios are still minor (e.g., $>1$:3), i.e. the initial SMBH mass ratios may have been far more disproportionate at the start of the merger sequence. This is particularly interesting for two cases: (1) post-mergers and (2) late-stage mergers prior to nuclear coalescence. In the case of post-mergers, the initial stellar mass ratios will presumably have been erased after the stellar nuclei merge and the system relaxes; here, one may not be able to trace the merger history for the SMBH pair at the center, and thus remnants of minor mergers could be mistaken for relaxed remnants of major mergers\footnote{Trainwreck merger remnants may offer the only clear, but indirect clue about the progenitor mass ratio.}. For late-stage mergers with separations $0.1\,\rm{kpc}<r_p<10\,\rm{kpc}$ (where the nuclei have not yet coalesced), the mass ratios of the SMBHs may not trace the mass ratios of the stellar nuclei if the stellar mass growth has not kept pace with the SMBH growth \citep[e.g.,][]{callegari2011}, producing systems where the secondary SMBHs appear overmassive relative to their host. Such a scenario could at least partially explain the overmassive secondary SMBH in Was 49b \citep{secrest2017}, which has a mass of $\sim10^8$~$M_\sun$ despite residing in a host with an apparent stellar mass of only $\sim6\times10^9$~$M_\sun$ in a $\sim1$:7 to $\sim1$:15 minor merger with the larger Was~49a. It would be interesting to compare the stellar mass ratios with the mass ratios of the SMBHs (if these values could be robustly determined) for all dual AGNs in minor mergers to gather evidence of whether the secondary SMBHs are commonly found to be overmassive like in the case of Was 49b, though such a comparison is beyond the scope of this work.

Numerical simulations suggest, however, that dual AGN observability timescales (above luminosity thresholds of $10^{43}$ erg s$^{-1}$) in minor mergers are short relative to major mergers \citep[$\sim$20--70 Myr vs.\ $\sim$100--160 Myr;][]{vanwassenhove2012,capelo2017}, and these timescales drop by up to a factor of $\sim4$ or more when taking into account realistic observability constraints on imaging (separations ranging from 1--10 kpc) and spectroscopic (velocity differences $\lesssim\,150$\,km\,s$^{-1}$) surveys \citep[see][]{vanwassenhove2012,capelo2017}. Dual AGNs in minor mergers may therefore not necessarily be observed as frequently as those in major mergers at any one point in time, but at the very least dual AGNs in minor mergers should be signposts for incredibly efficient merger-induced SMBH growth. Moreover, given (1) how much more frequently minor mergers occur relative to major mergers, and (2) the amount of growth the SMBHs undergo, particularly the SMBH in the less massive progenitor, it would seem plausible that -- averaged across cosmic time -- minor mergers could represent the dominant formation and/or growth pathway for dual AGNs, despite the shorter duty cycle for the AGNs. While beyond the scope of this work, it may be possible to explore this hypothesis using current cosmological simulations, and it remains important to exhaustively analyze these types of dual AGNs when  detected so that we may better understand this under-sampled population.

The fact also remains that minor mergers should lead to efficient pairings of SMBHs \citep{callegari2009,callegari2011,khan2012b,capelo2015}; at higher redshifts where mergers occur more frequently and gas fractions are higher \citep[the simulations of][are in fact designed to emulate higher redshift minor mergers]{callegari2009,callegari2011,vanwassenhove2012,capelo2015,capelo2017}, the coalescence of these binaries would be detectable in the future by the Laser Interferometer Space Antennae \citep[LISA,][]{amaro-seoane2017}. Intriguingly, simulations predict long timescales for the largest mass ratios \citep[i.e., 1:10,][]{callegari2011,capelo2015}; it is conceivable that such timescales at earlier epochs would increase the chances of minor mergers overtaking one another, leading to the formation of bound SMBH triplets or multiplets \cite[see][for a local redshift example of a kpc-scale triple AGN in a late-stage merger]{pfeifle2019b,liu2019}. This is particularly important, because the gravitational interactions between the SMBHs within a bound SMBH multiplet can shorten the timescales for coalescence of the inner binary \cite[e.g.,][]{ryu2018}, result in ejected SMBHs via gravitational slingshots \citep[e.g.,][]{hoffman2007,bonetti2018,bonetti2019}, and drive high orbital eccentricities that would affect the gravitational waveforms observable with LISA \citep[e.g.,][]{bonetti2019}.

\subsection{The Lack of Type~1/Type~1 Dual AGNs at Local $z$}
We know of Type~1/Type~1 AGN pairs and/or quasar pairs at higher redshift (beyond the local universe, $z>0.1$) in both earlier-stage mergers \citep[e.g.,][but some of these may be co-spatial rather than merger-induced]{brotherton1999,schneider2000,gregg2002,hennawi2006,hennawi2010,shalyapin2018,more2016,mcgreer2016,green2010} and late-stage mergers \citep[e.g.,][]{silverman2020}, yet apparently none have been found in early- or late-stage mergers in the local universe. This begs the question that we must be missing this potential population in the local universe. 

Given the prevalence of high absorbing columns in many confirmed dual AGNs in the local universe \citep[e.g.,][]{komossa2003,bianchi2008,pfeifle2019b}, as expected based on recent hydrodynamic simulations \citep[e.g.,][]{capelo2017,blecha2018}, one may expect to detect Type~2/Type~2 and Type~1/Type~2 pairs far more frequently than Type~1/Type~1 pairs, in spite of the fact that one might naively expect Type~1/Type~1 pairs to be more easily identified and confirmed via broad emission lines. Unfortunately, spectroscopic surveys such as the SDSS already suffer from biases against observing close pairs due to the fiber collision limit on the spectroscopic plates, and therefore one reason we may not have already identified a significant fraction of the Type~1/Type~1 pairs presumed to exist in the local universe is simply a lack of spectroscopic completeness for close pairs of AGNs. Couple spectroscopic incompleteness with the expectation that Type~1/Type~1 pairs should be intrinsically infrequent relative to other Seyfert pairings, and we can begin to see why such pairs have yet to be discovered locally. Higher redshift surveys typically rely upon photometric selection first, which does not suffer from a fiber collision bias, and so these surveys can more easily identify closely separated AGNs, but with two major caveats: (1) the imaging must be of sufficiently high angular resolution \citep[like Suburu/Hyper Suprime-Cam, e.g.,][]{silverman2020}, and (2) such a selection strategy (necessarily) biases one against weaker AGNs at higher redshifts and biases toward identifying the brightest pairs of AGNs and/or the least obscured AGNs. No equivalent search has been performed for dual AGNs in the local universe; all \textit{systematic} searches for dual AGNs in the optical band in the local universe have required spectroscopic redshifts and/or spectroscopic emission lines as a primary selection criterion, rather than arising from follow-up observations. Such a task would be no small feat, however: the angular resolution of facilities such as SDSS would preclude the unambiguous detection of photometric pairs in the latest-stage mergers, and a blind photometric search may be temporally prohibitive, thus enforcing a bias towards seeking out only the brightest photometric pairs at somewhat larger separations than the commonly studied late-stage mergers. 

\section{Conclusion}
\label{sec:conclusion}
Dual AGNs are predicted to be less commonly found in minor mergers than in major mergers, yet minor mergers can lead to substantial growth of the SMBH in the less massive companion and efficient pairings of bound SMBHs following the merger evolution. Here we studied two systems, J1421+4747 and J1713+3256, that appeared to host the first local redshift dual Type~1 AGNs in apparent minor mergers based upon SDSS spectroscopic measurements of both sources in each system. However, follow-up LBT spectroscopic observations have shed a final light on these systems, and we summarize this work here:
\begin{itemize}
    \item Fitting the SDSS spectra with BADASS, we found evidence for broad H$\alpha$ lines in the primary and secondary nuclei of both J1421+4747 and J1713+3256. We found FWHM values for the broad H$\alpha$ lines in the primary and secondary nuclei of $6073^{+24}_{-46}$ km~s$^{-1}$ and $4290^{+119}_{-118}$ km~s$^{-1}$ in J1421+4747 and $4649^{+55}_{-84}$ km~s$^{-1}$ and $3560^{+318}_{-190}$ km~s$^{-1}$ in J1713+3256, which are inconsistent at the $>$2--3$\sigma$ level, suggesting two distinct kinematic regions and hence two broad line regions in each apparent merger. In addition, each nucleus presented Seyfert-like optical spectroscopic narrow line ratios.
    \item Our new LBT optical spectroscopic observations have revealed, however, that while these systems consist of broad line AGNs in the primary nuclei, the secondary sources are actually foreground stars with characteristic rest-frame stellar absorption features (Ca triplet and Na D absorption lines in J1421+4747; Ca triplet, Ca K+H, and Na D absorption lines in J1713+3256) in the spatially resolved spectra. The previously identified broad and narrow emission lines observed in the SDSS spectra (and the LBT spectra, in the case of J1421+4747) of the secondary sources was likely the result of spillover light from the primary source, likely induced by the atmospheric seeing conditions during the previous SDSS observations. 
    \item Spatially resolved spectroscopic follow-up observations represent an incredibly important check for dual AGN studies that rely upon spectroscopic fiber measurements, particularly when objects are selected using fiber spectra at small separations, as was the case in this work. Sensitivity and wavelength coverage also played a critical role in this work: the presence of foreground stars may not have been clear if instruments with more limited wavelength ranges or lower sensitivity had been used here. Two of the three lines (the Ca triplet and Ca K+H absorption lines) that clinched the nature of the secondary objects reside at the extreme red and blue ends of optical spectroscopy, while high sensitivity was needed to identify both the Na D lines and the Ca triplets.
\end{itemize}

Confirmation of one or both of the systems studied here was an exciting prospect at the onset of this work, as it would have immediately increased the known number of dual AGNs hosted in minor mergers by $\sim$30\%-40\% while at the same time identifying dual AGNs with remarkably different optical spectroscopic characteristics---i.e. optically unobscured spectra, implying a lack of line-of-sight obscuration---than the currently known dual AGNs host by minor mergers and the known dual AGN population at large. Though this work did not identify new dual Type~1/Type~1 AGNs in minor mergers, when placed into the context of the theoretical works in the literature, it does expose some inadequacies in our understanding of dual AGNs (or AGNs in general) in minor mergers. For example, it is not clear whether minor merger simulations that have accounted for gas absorption \citep[e.g.,][]{capelo2017} find that minor mergers should produce preferentially obscured nuclei like simulations of major mergers \citep[e.g.,][]{capelo2017,blecha2018}. Hydrodynamic and radiative transfer simulations performed in \citet{blecha2018} predict that dual AGNs should be heavily obscured in major mergers and emitting strongly in the mid-IR as the intrinsic AGN continuum is reprocessed by the dust, but little focus was placed on minor mergers in that work; only a single 1:4.5 minor merger was studied in detail, and very little AGN activity (single or dual AGN) was found to arise, in contrast with previous works \citep[e.g.,][]{callegari2009,callegari2011,vanwassenhove2012,capelo2015,capelo2017}. Given the lack of attention given to minor mergers in simulations where both dust and gas attenuation is taken into account, there still remains a fundamental gap in our understanding of dual AGN mid-IR colors, fueling habits, and line-of-sight nuclear column densities in not just the more commonly examined minor mergers of 1:4-1:10 mass ratios, but also with regard to more severely unequal mass mergers with mass ratios of 1:15, 1:25, and beyond 1:30. Suites of hydrodynamic simulations that can account for line-of-sight absorption and dust attenuation down to tens of parsecs in resolution for a variety of minor mergers may have an important impact on our understanding of selection techniques and biases when seeking out these minor merger dual AGNs. 

\begin{acknowledgements}
We would like to thank the anonymous referee for their prompt and thoughtful review that helped to improve this work. R.W.P. and J.M.C. acknowledge support for this work through appointments to the NASA Postdoctoral Program at Goddard Space Flight Center, administered by ORAU through a contract with NASA. R.W.P would like to thank N. Latouf for support during this work. The authors would also like to thank Jennifer Power for carefully executing the LBT/MODS observations, and R.T. Gatto for useful discussions and assistance in the observations.

Funding for SDSS-III has been provided by the Alfred P. Sloan Foundation, the Participating Institutions, the National Science Foundation, and the U.S. Department of Energy Office of Science. The SDSS-III web site is http://www.sdss3.org/.

SDSS-III is managed by the Astrophysical Research Consortium for the Participating Institutions of the SDSS-III Collaboration including the University of Arizona, the Brazilian Participation Group, Brookhaven National Laboratory, Carnegie Mellon University, University of Florida, the French Participation Group, the German Participation Group, Harvard University, the Instituto de Astrofisica de Canarias, the Michigan State/Notre Dame/JINA Participation Group, Johns Hopkins University, Lawrence Berkeley National Laboratory, Max Planck Institute for Astrophysics, Max Planck Institute for Extraterrestrial Physics, New Mexico State University, New York University, Ohio State University, Pennsylvania State University, University of Portsmouth, Princeton University, the Spanish Participation Group, University of Tokyo, University of Utah, Vanderbilt University, University of Virginia, University of Washington, and Yale University.

The LBT is an international collaboration among institutions in the United States, Italy and Germany. LBT Corporation partners are: The University of Arizona on behalf of the Arizona Board of Regents; Istituto Nazionale di Astrofisica, Italy; LBT Beteiligungsgesellschaft, Germany, representing the Max-Planck Society, The Leibniz Institute for Astrophysics Potsdam, and Heidelberg University; The Ohio State University, and The Research Corporation, on behalf of The University of Notre Dame, University of Minnesota and University of Virginia.
\end{acknowledgements}

\facilities{WISE, SDSS, HST, LBT}

\software{APLpy \citep{robitaille2012}, pandas \citep{mckinney2010}, NumPy \citep{oliphant2006,walt2011,harris2020array}, SciPy \citep{virtanen2020}, Matplotlib \citep{hunter2007}, \textsc{ds9} \citep{joyce2003}, BADASS \citep{sexton2021}, Astropy \citep{2013A&A...558A..33A,2018AJ....156..123A}}

\appendix
\section{Other Similar Targets Drawn From SDSS }
\label{appendixA}
Here we provide the full list of 13 systems drawn from SDSS that show similar dual broad line classifications, based on our selection methodology described in Section~\ref{sec:intro}, which includes J1421+4747 and J1713+3256 (examined here), along with brief notes on some of the objects, most of which are likely the result of fiber spillover (as was observed in this work) and/or substantial contributions from both sources as a result of the fiber positions.

\citet{hou2019} recently examined Chandra X-ray imaging of J0805+2818 and found that only one of the nuclei are detected. \citet{husemann2020} recently examined J0805+2818, J0858+1822, J0947+6339, J1609+2830 in a study examining fiber spillover contamination in closely-separated dual AGN candidates selected using SDSS spectroscopic fiber measurements. In all cases but J1609+2830, they found that fiber spillover was the likely culprit for the double optical AGN signal. The X-ray results from \citet{hou2019} and optical spectroscopic results from \citet{husemann2020} are consistent with J0805+2818 hosting only a single AGN. J1609+2830 was recently observed with Keck, but the K-band observations show broad Pa$\alpha$ and broad H$_2$ emission lines in only the eastern nucleus. The weaker, western nucleus shows only H$_2$ emission and does not show any Pa$\alpha$ emission (Bohn, private communication). This is not to say that the Keck observation of J1609+2830 rules out a dual AGN scenario, but taken together with the results of \citet{husemann2020}, it is more likely that this system contains only a single AGN. 

J1421+4747 and J1713+3256, discussed in this paper, are unfortunately star-quasar pairs caught in chance projections. J1429+4447 does not show an obvious companion nucleus in the SDSS imaging. We checked the available HST data, but again we do not find any evidence for a secondary nucleus, so this object is very likely a single AGN. As far as we have found, J0322+0054, J1005+3414, and J1659+2446 have not been followed-up in any dual AGN study in the literature. J0222-0857 was nearly included in a study aimed at investigating AGN cross ionization \citep{keel2019}, but the target was unfortunately not observed during their follow-up long slit observations. The primary and secondary optical spectra in J0222-0857, J0322+0054, J1005+3414, J1429+4447, and J1659+2446 were obtained from fiber positions that unfortunately overlap in each case, and therefore both sets of spectra in each of these systems are expected to include substantial contributions from both sources.

Was 49b (J1214+2931) is a known Type~2/Type~2 dual AGN \citep{bothun1989,secrest2017}; the brighter of the two AGNs in this minor merger system is actually the offset AGN. Interestingly, the optical spectrum of the offset AGN superficially appears as a Type~1 AGN, but this is due to broad line emission back-scattered into the narrow line region. J1558+2723 is a little-known dual AGN in the Abell 2142 cluster confirmed via Chandra X-ray imaging by \citet{eckert2017}.

\begin{table}[ht]
\begin{center}
\caption{``Broad-line'' Pairs Drawn From SDSS}
\label{table:Otherspillovertargets}
\begin{tabular}{ccccc}
\hline
\hline
\noalign{\smallskip}
\noalign{\smallskip}
Target & RA & Dec & Redshift & Sep \\
(1) & (2) & (3) & (4) & (5)\\
\noalign{\smallskip}
\noalign{\smallskip}
\hline
\noalign{\smallskip}
J0222-0857 & 35.60884 & -8.95036 & 0.16663 & $6.1$ \\
           & 35.60928 & -8.95076 & 0.16636 &  \\
J0322+0054 & 50.61354 & 0.900561 & 0.16401 & $8.4$ \\
           & 50.61312 & 0.901283 & 0.16394 &  \\
J0805+2818 & 121.3471 & 28.3044  & 0.12843 & $5.3$ \\
           & 121.3475 & 28.30391 & 0.12864 &  \\
J0858+1822 & 134.6564 & 18.37266 & 0.05874 & $3.2$ \\
           & 134.657  & 18.37315 & 0.05893 &  \\
J0947+6339 & 146.9232 & 63.66148 & 0.13973 & $5.2$ \\
           & 146.9232 & 63.66089 & 0.13904 &  \\
J1005+3414 & 151.2828 & 34.24005 & 0.16184 & $6.1$ \\
           & 151.2836 & 34.24004 & 0.16206 &  \\
J1214+2931 & 183.5741 & 29.52869 & 0.06342 & $8.4$ \\
           & 183.5761 & 29.52965 & 0.06326 &  \\
J1421+4747 & 215.3752 & 47.79128 & 0.07327 & $7.0$ \\
           & 215.3739 & 47.79016 & 0.07260 &  \\
J1429+4447 & 217.3917 & 44.79740 & 0.20774 & $8.3$ \\
           & 217.3923 & 44.79692 & 0.20749 &  \\
J1558+2723 & 239.7103 & 27.39007 & 0.09352 & $6.9$ \\
           & 239.7097 & 27.39105 & 0.09517 &  \\
J1609+2830 & 242.3884 & 28.51613 & 0.17039 & $7.7$ \\
           & 242.3892 & 28.51622 & 0.16959 &  \\
J1659+2446 & 254.9156 & 24.77131 & 0.13921 & $5.0$ \\
           & 254.9162 & 24.77165 & 0.13824 &  \\
J1713+3256 & 258.3455 & 32.94123 & 0.10140 & $7.9$ \\
           & 258.3441 & 32.94111 & 0.10157 &  \\
\noalign{\smallskip}
\hline
\end{tabular}
\end{center}
\tablecomments{Other ``broad line'' pairs selected via SDSS using the selection criteria in Section~\ref{sec:intro}. Col 1: truncated merger designation. Col 2-4: right ascension, declination, and redshift of nuclei. Col 5: separation of the nuclei in kpc. Only two of these systems have been confirmed as dual AGNs: J1214+2931 (Was 49b) and J1558+2723; the remainder are likely fiber spillover or cross-ionization targets.}
\end{table}

\bibliography{sample631}{}
\bibliographystyle{aasjournal}



\end{document}